\documentclass{egpubl}
\usepackage{eg2025}
 
\ShortPresentation      
\CGFStandardLicense

\usepackage[T1]{fontenc}
\usepackage{dfadobe}  

\biberVersion
\BibtexOrBiblatex
\usepackage[backend=biber,bibstyle=EG,citestyle=alphabetic,backref=true]{biblatex} 
\electronicVersion
\PrintedOrElectronic

\ifpdf \usepackage[pdftex]{graphicx} \pdfcompresslevel=9
\else \usepackage[dvips]{graphicx} \fi

\usepackage{egweblnk} 

\usepackage{amsmath}
\usepackage{amssymb}

\usepackage{bm}

\title[3D Gabor Splatting]%
      {3D Gabor Splatting: Reconstruction of High-frequency Surface Texture using Gabor Noise}

\author[Haato Watanabe \& Kenji Tojo \& Nobuyuki Umetani]
{\parbox{\textwidth}{\centering Haato Watanabe$^{}$\orcid{0009-0006-0097-9310}, Kenji Tojo$^{}$\orcid{0000-0001-9415-0701} and Nobuyuki Umetani$^{}$\orcid{0000-0003-1251-970X} 
        }
        \\
{\parbox{\textwidth}{\centering $^{}$The University of Tokyo\\
       }
}
}

\begin{document}

\teaser{
 \includegraphics[width=176mm]{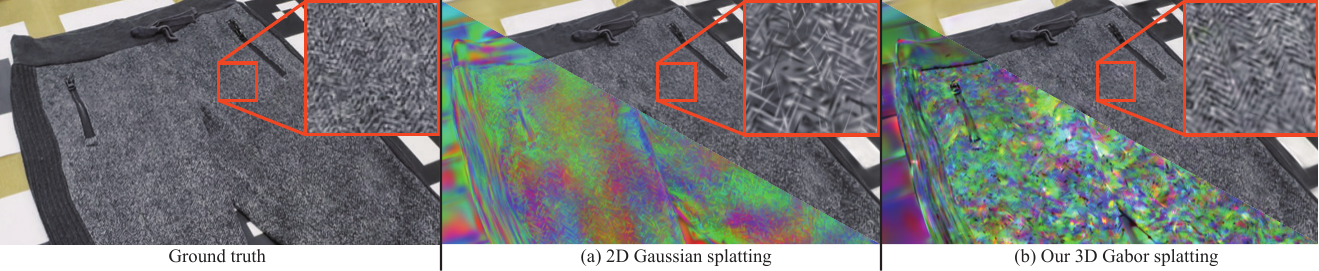}
 \centering
 \caption{While the 2D Gaussian splatting~\cite{huang20242d} generate splats with very high aspect ratios poorly reproducing the high-frequency texture in the target object (a), our 3D Gabor splatting approximates the texture better (b) in this \textsc{Sweat} dataset. The left-bottom half of each image shows the splats' shapes visualized with random colors.
 The number of the kernels is 3.5k for both models.} 
\label{fig:teaser}
}

\maketitle
\begin{abstract}
3D Gaussian splatting has experienced explosive popularity in the past few years in the field of novel view synthesis.
The lightweight and differentiable representation of the radiance field using the Gaussian enables rapid and high-quality reconstruction and fast rendering.
However, reconstructing objects with high-frequency surface textures (e.g., fine stripes) requires many skinny Gaussian kernels because each Gaussian represents only one color if viewed from one direction.
Thus, reconstructing the stripes pattern, for example, requires Gaussians for at least the number of stripes. 
We present 3D Gabor splatting, which augments the Gaussian kernel to represent spatially high-frequency signals using Gabor noise.
The Gabor kernel is a combination of a Gaussian term and spatially fluctuating wave functions, making it suitable for representing spatial high-frequency texture.
We demonstrate that our 3D Gabor splatting can reconstruct various high-frequency textures on the objects.

\ccsdesc[500]{Computing methodologies~Rasterization}
\ccsdesc[500]{Computing methodologies~Reconstruction}

\printccsdesc   
\end{abstract}

\section{Introduction}
%
The 3D Gaussian splatting (3DGS)~\cite{kerbl3Dgaussians} has ushered in a new era for novel view synthesis and 3D object reconstruction.
The 3DGS represents a scene with a set of Gaussian primitives, which can be efficiently rendered by leveraging the rasterization procedure.
In addition, the post processes, such as object removal after reconstruction, are straightforward in this representation.
Thanks to the differentiability of the 3DGS, the training converges fast and faithfully reconstructs challenging scenes such as vegetation once the training converges. 
%


%
However, reconstructing objects with high-frequency surface textures seen in many textiles needs many primitives to reproduce the high color variance in the pattern, although the opacity variance is low. 
The problem stems from the limited color variation in one primitive.
We present 3D Gabor splatting, which can efficiently express high-frequency detailed texture by enriching the color variation of 3DGS. 
Our research was inspired by Gabor noise~\cite{lagae2009procedural}, a procedural texture generation method using the Gabor kernel, consisting of the Gaussian and sinusoidal wave function terms.
By using the 2D Gaussian splatting (2DGS)~\cite{huang20242d} formulation, our method augments 3DGS primitives with the Gabor kernel to enable each primitive to approximate high-frequency color variation.
We demonstrate that our 3D Gabor splatting can reproduce various high-frequency textures. 

\section{Related Work}

\paragraph*{View synthesis} 
Snavely et al.~\cite{agarwal2011building} proposed the structure from motion (SfM) technique that estimates each camera position and direction and reconstructs a sparse point cloud of the scene given a set of photos taken from multiple view angles.
Mildenhall et al.~\cite{mildenhall2020nerf} presented neural radiance fields (NeRF) that reconstruct a radiance field of the 3D scene using implicit neural representation. 
However, every sampling of color in space requires neural network inference, which makes rendering slow, and post-processing after reconstruction is difficult.



\paragraph*{Gaussian splatting}
Kerbl et al.~\cite{kerbl3Dgaussians} present the seminal work of 3D Gaussian splatting (3DGS), which utilizes a set of Gaussian primitives to represent a radiance field.
Optimizing each position, covariance, color, and opacity enables the approximation of the object surface using Gaussian primitives. 
The initial position of primitives is set on cloud points that are retrieved using structure from motion~(SfM)~\cite{agarwal2011building}. 
Gaussian splatting converges faster and typically produces higher-quality output than NeRF. 
Real-time rendering can also be conducted.
A significant amount of research has been aimed at extending Gaussian splatting. Please refer to the survey~\cite{wu2024recent} for a comprehensive overview.
Here, we focus on enriching primitive representation ability to reconstruct high-frequency texture surface. 

Huang et al.~\cite{huang20242d} present the 2D Gaussian splatting (2DGS), which is an improvement of 3DGS assuming that the Gaussian primitives are flat the object normal direction (i.e., Gaussian distribution on a 3D plane).
The 2DGS can reconstruct the flat surface of objects more precisely than the original 3DGS.
However, both 3DGS and 2DGS assume uniform color inside a primitive, thus incapable of representing color variation inside a primitive. 
Nevertheless, we leverage the flat geometry of the 2DGS to incorporate the Gabor kernel into the Gaussian primitives.

\paragraph*{Gabor noise}
A Gabor kernel consists of 2D Gaussian distribution and sinusoidal wave (i.e., harmonic) terms. 
Lagae et al.~\cite{lagae2009procedural,lagae2011filtering} proposed Gabor noise a procedural texture generation method using Gabor kernel. 
%
%
Galerne et al.~\cite{galerne2012gabor} proposed a method that can estimate parameters of Gabor noise from an image, using the difference of power spectrums between the example and Gabor noise. 
Jeschke et al.~\cite{jeschke2011curve} further applied Gabor noise to the diffusion curve to enhance the expressiveness.
Our research is not a 2D texture synthesis method; instead, we augment the Gaussian primitives by adding the harmonic term in the Gabor kernel that allows the reconstruction of the texture on the input objects.
The Gabor splatting~\cite{wurster2024gabor} is a method for representing 2D images using Gabor primitives. 
While their method can represent high-resolution images, they focus on 2D image representation, and while we aim at reconstructing 3D scenes.

\section{Background: 2D Gaussian Splatting}
\label{sec:background}

For self-consistency, this section briefly explains the existing 2D Gaussian splatting~\cite{huang20242d} on which our algorithm is based.
2DGS assumes that the width of the Gaussian in a normal direction is zero.
Thus, the density is distributed inside a 3D plane, which can be parameterized with the \textit{local} 2D coordinate $(u,v)\in\mathbb{R}^2$ as
\begin{equation}
    \vec{p}(u,v) = \vec{q} + u s_u \vec{t}_u + v s_v \vec{t}_v,
\end{equation}
where $\vec{q}\in\mathbb{R}^3$ is the center position of the Gaussian kernel, $s_u$ and $s_v$ are the scaling factors in the principle directions and $\vec{t}_u\in\mathbb{R}^3$ and $\vec{t}_v\in\mathbb{R}^3$ are the unit vectors in the direction of principle axes. 
This 2D local coordinate allows us to formulate 2D wave functions in the 3D Gaussian kernel.
To find the local coordinate $(u,v)$ at a pixel, the 2D Gaussian splatting analytically computes the intersection between the plane and the viewing ray from the pixel.
The Gaussian kernel of a primitive
\begin{equation}
    \mathcal{G}(u,v) = \exp\left\{- \frac{u^2+v^2}{2} \right\},
\end{equation}
defines the spatial distribution of the alpha channel while $\alpha\in\mathbb{R}$ define the overall opacity.
Specifically, the 2DGS~\cite{huang20242d} slightly modify $\mathcal{G}$ to enhance stability in edge cases as $\hat{\mathcal{G}}$.
Here, for brevity, we refer to the original paper for  $\hat{\mathcal{G}}$.
Given the alpha channel value $\alpha\;\hat{\mathcal{G}}(u,v)$ for all the primitives, the 2DGS computes the final color of a pixel by the alpha blending algorithm.
While the color of a primitive is homogeneous in 2DGS, we spatially change the color inside a kernel to enhance the detail representation.

\begin{figure}[t!]
    \centering
    \includegraphics[width=84mm]{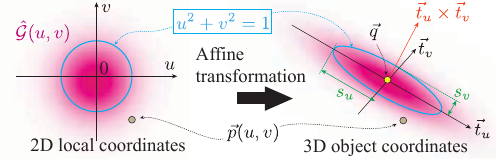}
    \caption{Construction of the 2D Gaussian splatting~\cite{huang20242d}.}
    \label{fig:2dgs}
    \vspace{-5mm}
\end{figure}

\section{Method}
\label{sec:method}

\paragraph*{Multiple wave formulation} The original Gabor kernel is the multiplication of a Gaussian function and a single sinusoidal wave function (i.e., harmonic term). 
However, fitting both the frequency and orientation of the wave functions in a Gabor kernel into high-frequency input images is a highly nonlinear problem. 
Thus, a na\"ive gradient-based optimization is typically stuck at a local minimum and fails to converge.
To alleviate this problem, we introduce a simple remedy in which one kernel has a weighted sum of the \textit{multiple} wave functions, and we optimize only the frequencies of the wave functions while fixing their orientations.
The orientations are \textit{uniformly sampled} from the circumference.
The insight behind this procedure is that by combining the multiple wave orientations, some wave orientations capture the principal wave signal in the input image.

While \cite{wurster2024gabor} also uses multiple wave functions, their formulation differs significantly from ours.
They fix both the frequencies and orientations of the wave functions, and all the orientations are in the same horizontal directions in the local coordinates.
Our formulation enjoyed more representation capability while removing the nonlinearity stem from the orientation optimization. 
Note that our formulation is inspired by the 2D hair orientation map computation~\cite{paris2004capture}, where the orientations of the wave functions in the Gabor kernel are predefined and uniformly sampled.

%

\paragraph*{Kernel formulation}
Using the 2DGS representation~\cite{huang20242d}, we incorporated our Gabor kernel with multiple wave functions into the 3D Gaussian splatting.
While the alpha value for each Gabor primitive is defined the same as the 2DGS (i.e., $\hat{\mathcal{G}}(u,v)$ in Sec.~\ref{sec:background}), we enrich the spatial color distribution by several wave functions.
The $i$-th wave function of the kernel orients toward direction $i\pi/N$ in the polar coordinate where $N\in\mathbb{Z}^+$ is the number of orientation samples.
The choice of the sampling number $N$ is a trade-off between expressiveness and calculation cost of training and rendering.
We currently set $N=4$ (i.e., samples at $45^\circ$ interval).
This sampling can be sparse because the primitives have the ability to fine-tune their orientation by changing their principal directions.
The phase function $\theta_i$ for $i$-th wave function becomes
\begin{eqnarray}
\label{eqn:phase}
    \theta_i(u,v) &=& 2 \pi f_i \left(\cos \frac{i\pi}{N}, \sin \frac{i\pi}{N} \right)(u,v)^T + \phi_i,
\end{eqnarray}
where the $\phi_i\in\mathbb{R}$ is the phase shift function inspired by~ \cite{lagae2011filtering, galerne2012gabor}.
The $f_i\in\mathbb{R}$ is the spatial frequency in the local coordinate (not in the coordinates of the captured 3D object). 
In other words, the frequency $f_i$ roughly describes how many wave cycles in a single primitive.

\begin{figure}[htbp!]
    \centering
    \includegraphics[width=84mm]{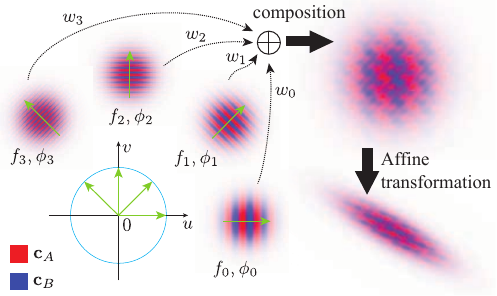}
    \caption{Construction of our Gabor kernel. The multiple wave functions, whose directions are uniformly sampled, are composed in the local coordinate of the kernel. Then, affine transformation computes the color distribution in the object space.}
    \label{fig:construction-gabor-kernel}
    \vspace{-3mm}
\end{figure}

Each wave function has a spatial distribution that interpolates two independent colors $\mathbf{c}_A\in\mathbb{R}^3$ and $\mathbf{c}_B\in\mathbb{R}^3$ that are defined for each kernel.
Finally, the color of each kernel at the local coordinates $(u,v)$ becomes
\begin{eqnarray}
\label{eqn:color}
    \mathbf{c}(u,v) = \sum_{i=0}^{N-1} w_i \left\{\mathbf{c}_A \frac{1 + \cos\theta_i}{2} + \mathbf{c}_B \frac{1 - \cos\theta_i}{2} \right\},
\end{eqnarray}
where the $w_i\in\mathbb{R}$ is the weight for each wave function
(see Fig.~\ref{fig:construction-gabor-kernel}).
Note that we currently ignore view-dependent color change as our current main target is the detailed texture of the diffuse surface (e.g., garment).
Given the alpha value $\hat{\mathcal{G}}(u,v)$ and the color $\mathbf{c}(u,v)$ given by \eqref{eqn:color}, we compute the alpha blending of the kernels similar to the 2DGS~\cite{huang20242d}.

\paragraph*{Optimization}
At the training time, we optimize the parameters of our Gabor kernel, which are $\vec{q}, s_u, s_v,\vec{t}_u,\vec{t}_v, \alpha, \mathbf{c}_A, \mathbf{c}_B$, and $\{w_i,\phi_i,f_i\}$ where $i\in \{0,\ldots,N-1\}$.
Similar to the original 3DGS~\cite{kerbl3Dgaussians}, we initialize the kernel position $\vec{p}$ as the points generated from the structure from motion~(SfM).
The remaining parameters are optimized from scratch from random initial values. 
We use the same loss function and the same number of training iterations (30k) as the 2DGS~\cite{huang20242d}. 
%

%

\section{Results}


\paragraph*{Dataset}
Our approach targets high-frequency texture on 3D objects.
Thus, for comparison, we collect four datasets (\textsc{Sweat}, \textsc{Vest}, \textsc{Shirt}, and \textsc{Boots}) that are photos of high-frequency textured garments and shoes.
Each dataset has at least 60 pictures that are in full HD (1920 x 1080 px) resolution taken by the authors using iPhone 15 Pro. 
We use COLMAP~\cite{schoenberger2016sfm} to create SfM point clouds and estimate camera parameters. 
We put fiducial markers on the background of the target objects to help the accurate estimation of camera parameters.

\paragraph*{Implementation}
We implement our 3D Gabor Splatting based on the published code of 2DGS~\cite{huang20242d}.
%
We will publish our code and the datasets.
To simplify the comparison condition, we turned off the view-dependent color from the spherical harmonics.
In addition, for comparison, we made the number of resulting primitives the same between ours and the 2DGS by turning off the densification.  
We conducted all experiments on a single GeForce RTX 3090 GPU.
The training takes roughly 30 min for 2DGS and 43 min for our 3D Gabor splatting. 
The runtime speed of our method is 65-95 frames per second (FPS) for all the models, whereas 130-176 FPS for the 2DGS.
%
Please see the supplemental video for more detail.



%



\begin{figure}[t!]
    \centering
    \includegraphics[width=84mm]{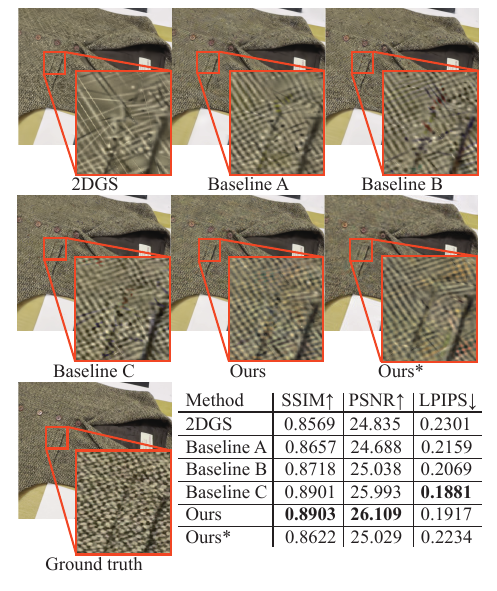}
    \caption{Qualitative and quantitative ablation study against three baselines. The single wave function inside a primitive (Baseline A) shows blurry output. The multiple wave functions in a single wave orientation (Baseline B) show false color. The effect of removing phase shift (Baseline C) is small in appearance, but the quantitative scores (right) slightly degrade. The Ours* is our result with approximately half the number of primitives, reducing the data size to the same as the 2DGS (both 2.411MB), while still showing improvements against the 2DGS.}
    \label{fig:ablation}
    \vspace{-2mm}
\end{figure}

\paragraph*{Quantitative evaluation} 
We used one-eighth of the dataset as the test dataset and the others for training. 
Similar to 2DGS~\cite{huang20242d}, we quantitatively compare the quality of the result by computing SSIM, PSNR, and LPIPS metrics. 
As shown in Table~\ref{tab:metrics} and Fig.~\ref{fig:ablation}, our method exhibits better scores in all three metrics than 2DGS.

\setlength{\tabcolsep}{4pt}
\begin{table}[b!]
\footnotesize
\begin{tabular}{l|ccc|ccc} 
    \multicolumn{1}{c|}{Dataset} & \multicolumn{3}{c|}{3D Gabor splatting} & \multicolumn{3}{c}{2D Gaussian splatting} \\ name & SSIM~$\uparrow$ & PSNR~$\uparrow$ & LPIPS~$\downarrow$ & SSIM~$\uparrow$ & PSNR~$\uparrow$ & LPIPS~$\downarrow$ \\ \hline
    \textsc{Sweat} & $\mathbf{0.872}$  & $\mathbf{25.69}$  & $\mathbf{0.232}$  & 0.852 & 25.15 & 0.276 \\
    \textsc{Boots}      & $\mathbf{0.867}$  & $\mathbf{23.59}$  & $\mathbf{0.305}$  & 0.853 & 23.23 & 0.329 \\ 
    \textsc{Shirt}      & $\mathbf{0.849}$  & $\mathbf{24.12}$  & $\mathbf{0.262}$  & 0.815 & 22.13 & 0.311 \\ 
\end{tabular}
\caption{Metrics scores of 3D Gabor splatting for \textsc{Sweat} in Fig.~\ref{fig:teaser} and  \textsc{Boots} and \textsc{Shirt} in Fig.~\ref{fig:convergence}.}
\label{tab:metrics}
\vspace{-3mm}
\end{table}





%

\paragraph*{Ablation study} 
%
%
Fig.~\ref{fig:ablation} compares our formulation against other simpler formulations on the \textsc{Vest} model as: (Baseline A) a na\"ive approach using a single wave function (i.e., $N=1$ in our method), (Baseline B) a kernel has a weighted sum of four harmonics fixed along the u-axis of the local coordinate (instead of our uniformly sampled directions), which is an approach roughly corresponds to the 3D version of the  Gabor Splatting~\cite{wurster2024gabor}, and (Baseline C) a variation of our method where the phase parameters $\phi_i$ are fixed to be zero.
Fig.~\ref{fig:ablation} shows our method achieved the highest SSIM and PSNR scores. 
We also demonstrate that our method achieves better accuracy than 2DGS, even with the same data size.

\begin{figure}[t!]
    \centering
    \includegraphics[width=84mm]{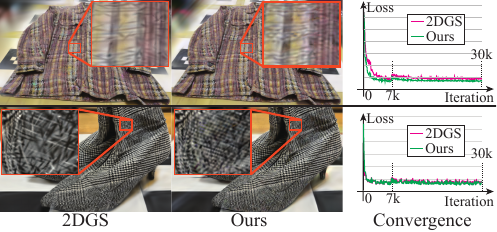}
    \caption{Comparison of the converged results using the 2D Gaussian splatting (2DGS)~\cite{huang20242d} (left) and our (middle) model. Our model converges faster to a smaller loss (right). The \textsc{Shirt} (top) dataset has 4.5k primitives, and the \textsc{Boots} (bottom) dataset has 4.2k primitives.}
    \label{fig:convergence}
    \vspace{-2mm}
\end{figure}

\paragraph*{Convergence}
Fig.~\ref{fig:convergence} shows the convergence of loss for our method and the 2DGS~\cite{huang20242d} on the \textsc{Boots} and \textsc{Shirt} dataset. 
Our method converges stably and faster than 2DGS, resulting in a more accurate representation of colorful and highly detailed textures.
Note that the sharp increase in loss value at the 7001st iteration is due to the introduction of the normal consistency loss~\cite{huang20242d}.
%

\section{Conclusion}
\label{sec:conclusion}
We present 3D Gabor splatting for high-frequency detailed texture reconstruction on 3D objects.
%
%
We demonstrated the effectiveness of our approach quantitatively and qualitatively through the comparison against 2DGS and ablation study.
%
%


\paragraph*{Future work} 
%
%
While we do not experience aliasing problems from the high-frequency wave function, filtering might be necessary in the future. 
We still need to extend our method to handle more complicated scenes, including view-dependent color change, using spherical harmonics.
%
%
Finally, we are interested in approximating sinusoidal functions in CUDA to accelerate rendering.

%
%

\vspace{-1mm}
\section*{Acknowledgements}
\label{sec:acknowledgements}
This work was financially supported by I. Meisters inc.
\vspace{-1mm}


\printbibliography                

\end{document}